\documentclass[12pt,a4paper]{article}
\pdfoutput=1
\usepackage{jheppub}
\usepackage[T1]{fontenc} 
\usepackage{epsfig} 
\usepackage{amscd}
\usepackage{verbatim}
\usepackage{latexsym}
\usepackage{amsfonts,amsthm,amsmath,mathtools}

\notoc 

\title{\centering Integral identities for 3d dualities with SP(2N) gauge groups}
\author[a]{Antonio Amariti}
\affiliation[a]{Laboratoire de Physique Th\'eorique de l'\'Ecole Normale Sup\'erieure \\
24 Rue Lhomond, Paris 75005, France}

         \emailAdd{amariti@phys.lpt.ens}

\abstract{In this note we study the reduction of 4d Seiberg duality 
to 3d for SP(2N) SQCD with an adjoint field. 
We follow a general prescription that consists in compactifying the dual 4d theories on the circle.
This generates an effective 3d duality. The pure 3d duality is obtained by combining the 
zero radius limit with a real mass flow.
Here we perform this limit by a double scaling procedure: we turn on  
real masses proportional to the radius before shrinking the circle.
We apply this mechanism to the reduction of the 4d superconformal index
to the three sphere partition function. 
While the reduction of the 4d index on the circle is straightforward, 
the 3d limit necessitates the double scaling. 
We describe this limit on the index, finding
the integral identity for the partition functions of the 3d dual theories. }

\begin{document}

\maketitle
\newpage
\tableofcontents

\section{Introduction}

In the last decade a new connection between physics and mathematics 
became attractive, especially after  \cite{Pestun:2007rz}. It is related to 
the application of localization techniques to the calculation of 
supersymmetric partition functions on compact manifolds.
This method is useful from the physical perspective because it
allows to capture quantum informations 
of the flat space theory from the knowledge of the exact partition function on the curved space.

This technique has applications in  strong/weak dualities  in field theory. 
For example the superconformal index can be used as a check  of
Seiberg duality.
The index corresponds to the partition function computed on  
$S^3 \times S^1$ and it is
a topological invariant quantity, counting a set of protected BPS multiplets,  respecting a shortening condition \cite{Kinney:2005ej,Romelsberger:2005eg}.
Matching the superconformal index between Seiberg dual models 
\cite{Dolan:2008qi,Spiridonov:2008zr,Spiridonov:2009za}
corresponds, in the mathematical literature, to study the 
transformation properties
of elliptic hypergeometric integrals \cite{Rains15,Spiridonov15}.
This connection between mathematics and physics has been subsequently extended to the 3d case. In this case the compact manifold is the squashed three sphere $S_\mathbf{b}^3$ \cite{Jafferis:2010un,Hama:2010av,Hama:2011ea}.
Matching the partition function of 3d dualities 
\cite{Willett:2011gp,Benini:2011mf}
corresponds to study the 
transformation properties of hyperbolic hypergeometric integrals \cite{VdB}.
Moreover, the 3d identities can be derived from the 4d ones by a limiting procedure 
\cite{Dolan:2011rp,Niarchos:2012ah,Gahramanov:gka,Aharony:2013dha,Amariti:2014iza}.
This procedure has a counterpart in the physical literature: by compactifying the 4d theories on 
a circle the 3d dualities can be derived from the 4d ones.
The generality of this procedure allows a further step in the link between mathematics and physics:
hypergeometric integral identities can be derived from elliptic integral identities with the
aid of a physical insight.
This program has been started in \cite{Aharony:2013dha} for SQCD like theories and extended in
\cite{Amariti:2014iza} to the case of $U(N)$ SQCD with adjoint matter.

These results are derived by following a stepwise reduction \cite{Aharony:2013dha}.
In the first step one starts from the 4d \emph{parent} electric and magnetic dual theories
and obtains a pair of  3d \emph{daughter} dual   theories. These theories are 
effective because they involve the  KK monopoles from the finite size of the circle.
This procedure prevents the generation of 4d anomalous symmetries in 3d: in the reduction
of the 4d index to the 3d partition function one can observe this effect in an extra constraint on the 
holomorphic parameters. The 3d identities are indeed valid only if the parameters of the theory 
satisfy some relation compatible with the superpotential interaction involving the 
KK monopoles.
The second step of the reduction consists of deriving the conventional 3d dualities: they 
are obtained by a further RG flow in  field theory. This flow allows to consistently 
shrink the radius of the circle to zero size preserving the duality. We will refer to this step as the 3d limit.
On the partition function this flow corresponds to a limit on the parameters involved in the 3d identities.

In this paper we derive the relation between the three sphere partition 
functions of $\mathcal{N}=2$ 3d dual gauge theories
with symplectic gauge group, fundamental and adjoint matter.
The relation is obtained by compactifying the identity between the superconformal 
indices of the 4d parent theories discussed in \cite{Leigh:1995qp}.
This 4d duality reduces to 3d to the duality of \cite{Kim:2013cma}.
We first compactify the theories on the circle 
obtaining the 3d effective duality as discussed above.
Then we flow to the 3d limit by considering some large real masses and by higgsing the dual gauge theory.
In this case some care is required: a double scaling limit on the real masses and on the radius of the circle is necessary.
This procedure has some consequences in the magnetic phase.
In this phase one has usually to map the vacua of the electric theory by higgsing the 
gauge group. This procedure generates an 
extra gauge sector. 
This sector can be alternatively described by a set of interacting singlets, namely the
mesons and the monopoles of the gauge theory.
By following this procedure we obtain the equality 
between the 3d partition functions of the 
dual phases of \cite{Kim:2013cma}.

The paper is organized as follows.
In section \ref{sec:review} we review some general aspects of the 4d/3d reduction at field theory level.
We also review the reduction in the D-brane engineering and discuss some general aspects of the 
reduction of the 4d index to the 3d partition function.
In section \ref{sec:reductionSP} we discuss the reduction for the case we are interested in: $SP(2N)$ gauge theory with fundamentals and adjoint matter.
In this case we reproduce the analysis of \cite{Amariti:2015yea,Amariti:2015mva} in terms of field theory, 
and, in the 3d limit, we obtain the duality discussed in \cite{Kim:2013cma}.
In section \ref{sec:partition} we obtain the main results: the relation between the partition function 
of the $SP(2N)$ theories with adjoint matter.
We obtain the expected identity mapping the partition functions in the dual phases.
In section \ref{sec:antisymmetric} we discuss the case with antisymmetric matter.
In section \ref{sec:conclusions} we conclude. In appendix \ref{sec:appendix} we give some 
mathematical definition concerning the 3d partition function on $S_\mathbf{b}^3$,
focusing on $SP(2N)$ SQCD with tensor matter.
In appendix \ref{doublesc} we study the double scaling limit on the 4d index.

\section{Review of the 4d/3d reduction}
\label{sec:review}
In this section we review some relevant aspects of the reduction of 
4d Seiberg duality to 3d.
In the first part  we illustrate the field theory prescription. 
In the second part  we describe the brane realization of this mechanism.
We conclude by discussing some general aspects of 
the reduction of the 4d index to the 3d partition function.

\subsection{Field theory}

Consider a pair of 4d $\mathcal{N}=1$ Seiberg dual gauge theories.
The 4d duality between the electric and magnetic \emph{parents} can be reduced to a duality between 3d \emph{daughter} theories by following the general procedure discussed in \cite{Aharony:2013dha}.

This procedure consists of first reducing the theories on $\mathbb{R}^3 \times S^1$,
and consider the circle at finite radius, $r$. The two 4d dual theories must be  
considered as effective 3d theories: in the IR there are non-perturbative effects,
related to monopoles acting as instantons in the compact Coulomb branch. 
There are two types of these monopoles, called  BPS and KK monopoles.
The first class of monopoles is typical in 3d theories. These monopoles 
act on the Coulomb branch, that is parameterized by the 
chiral multiplet $\Sigma= \sigma/g_3^2 + i \varphi$, where 
$\sigma$ is the real scalar in the vector multiplet, 
 and 
$d \varphi \simeq  * F$  is the dual photon. 
The 3d gauge coupling $g_3$ 
 is related to the 4d one by $g_4^2 = 2 \pi r g_3^2$.

The second class of monopoles, the KK monopoles, are due to the finite size  of the circle.
Their presence has to be considered if the
radius of the circle is non-vanishing.
The relation between the pair of 4d dual theories at energies lower than the radius is an effective duality. 
The presence of the KK monopoles is summarized in an Affleck-Harvey-Witten (AHW) \cite{Affleck:1982as} like superpotential, 
called $\eta$-superpotential, $W_\eta$.
For a gauge theory with Lie algebra $G$ the KK monopoles are related to the 
extra affine root of the algebra.
\\
In formulas we have \cite{Davies:1999uw,Davies:2000nw}
\begin{equation}
  \label{W-afffine}
  W(\Sigma) = W(\Sigma)_{\text{BPS}} + W(\Sigma)_{\text{KK}} \equiv
  \sum_{i=1}^{rk(G)} \frac{2}{\alpha_i^2} e^{\alpha_i^* \cdot \Sigma} +  \frac{2 \eta}{\alpha_0^2}
e^{\alpha_0^* \cdot \Sigma}
\end{equation}
where $\alpha_i$ are the simple roots of the Lie algebra and $\alpha_0$ corresponds to the extra affine root. The coroots are identified with a $*$.
 The constant $\eta$
is associated to the holomorphic scale of the 4d theory by 
$\eta=\Lambda^b $.
\\
The effective 3d duality can be further reduced to a more conventional one.
This reduction consists of sending $W_\eta$ to zero. This has to be consistently done in the electric and the magnetic phase.
In general this limit is consistent on both sides of the effective duality 
in presence of a real mass flow. In some cases there is also a not trivial mapping between the vacua of the electric and of the magnetic theory.
For example, by choosing the vacuum of the electric theory at the origin of the Coulomb branch,  one may have to choose 
the scalar $\sigma$ of the magnetic gauge group at non vanishing vev, 
corresponding to the topological vacuum of \cite{Intriligator:2013lca}.  
The vacuum breaks the gauge symmetry of the magnetic theory.
This breaking isolates an extra gauge sector that has
has been observed to be dual to a set of singlets \cite{Amariti:2015mva}. These singlets interact 
through an AHW superpotential with the monopoles of the magnetic gauge group. 
These singlets are naturally interpreted as the electric monopoles acting as elementary 
degrees of freedom in the magnetic dual theory.
At the end of this process the conventional
3d dualities of the type discovered in \cite{Aharony:1997gp,Karch:1997ux}
are obtained.

\subsection{Brane engineering}

In this section we review the 4d/3d reduction at brane level. This program started in \cite{Amariti:2015yea}
and it has been generalized in \cite{Amariti:2015mva}. See also \cite{Amariti:2015xna} for further extensions to s-confining theories
\footnote{The field theory reduction of 4d s-confining gauge theories has been treated in 
\cite{Csaki:2014cwa,Amariti:2015kha}.}.

Consider a type IIA brane setup describing an $\mathcal{N}=1$  4d gauge theory. The gauge theory lives on a stack of D4 branes
stretched between a pair of non parallel NS and NS' branes.
The fundamental flavors are introduced by  a stack of D6 branes.
Real gauge groups and tensor matter are associated to the addition of orientifold planes.

The 4d Seiberg duality is associated to the Hanany-Witten (HW) transition \cite{Hanany:1996ie}, i.e. the exchange of 
non parallel NS branes.
In this process a D4 brane is generated each time a D6 brane crosses 
a non-parallel NS brane. This brane creation mechanism
modifies the rank of the dual gauge group properly.

The duality is reduced to 3d by compactifying one space direction, along which all the branes are extended, for example  $x_3$, see (\ref{tableD}).
By T-dualizing along $x_3$ one obtains a type IIB brane setup. 
The radius $r$ of the circle becomes $\alpha'/r$ in the T-dual frame.
The 3d BPS monopoles are associated to D1 branes stretched along $x_3$ and $x_6$ between pairs of D3 and NS branes. Also the KK monopoles are captured by this construction, because the $x_3$ direction is compact \cite{Hanany:1996ie,deBoer:1997ka}. They correspond to the D1 brane stretched between the first and the last D3 brane on the circle.
Equivalently this construction can be  studied in the S-dual frame, where the role of the monopoles is played by  F1 strings \cite{Garland:1988bv,Hanany:2001iy}.
The type IIB brane setup and its dual configuration, obtained after the HW transition, represent the 3d effective duality discussed above.

\begin{figure}
\begin{center}
\includegraphics[width=15cm]{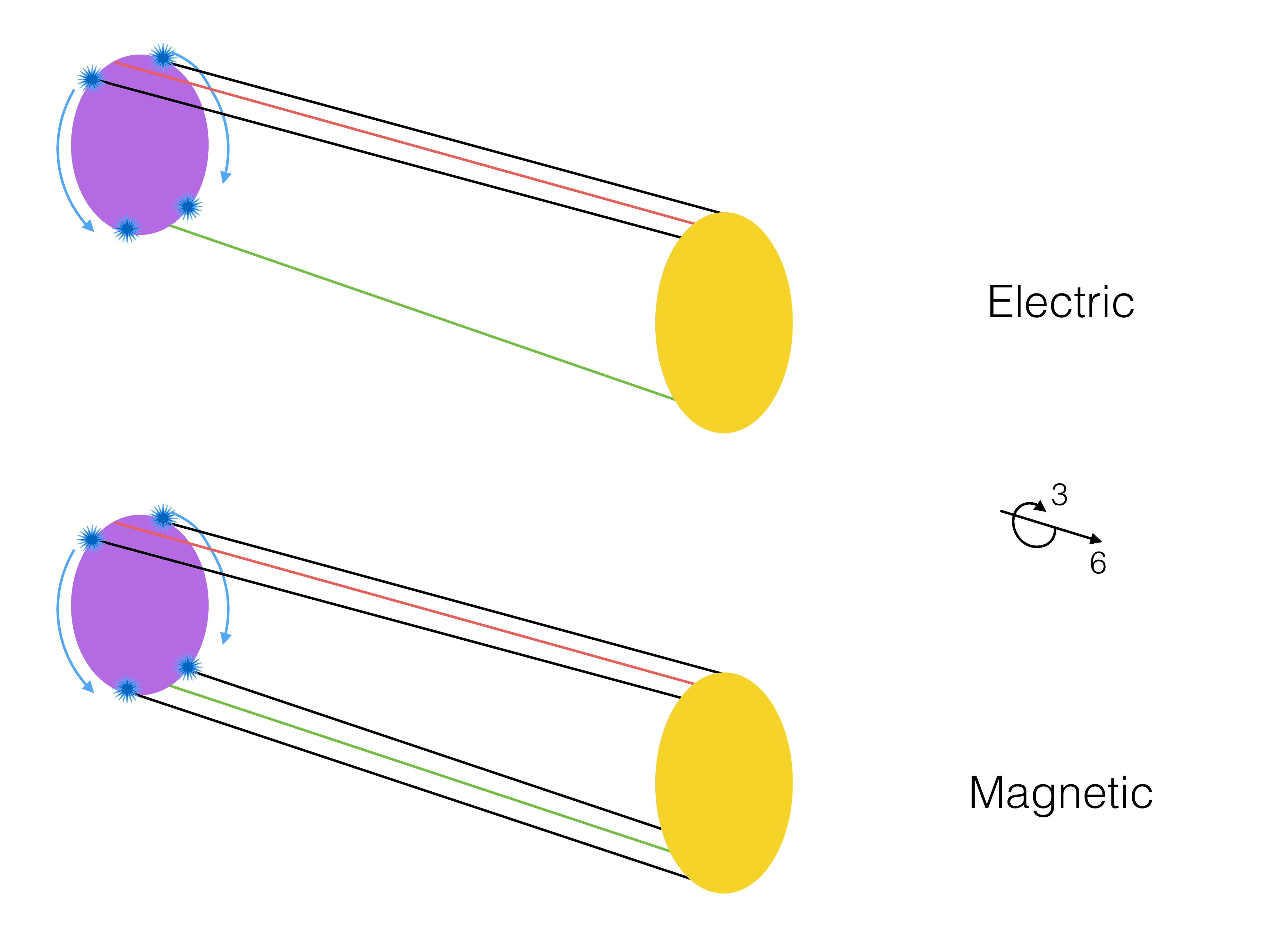}
\caption{Representation of the 3d limit in the IIB brane setup. The NS and NS' branes are represented in purple and yellow respectively. The D5 flavor branes are blue points in this picture and some of them are rotated along the compact direction $x_3$. The pair of orientifolds is associated to the pair of red and green lines, placed respectively at $x_3=0$ and $x_3=\pi \alpha'/r$. The D3 branes are represented by black lines: in the magnetic theory some of them are created by an HW transition also at  
$x_3=\pi \alpha'/r$.}
\label{O43}
\end{center}
\end{figure}

The second step of the analysis concerns the 3d limit. This limit in the
brane regime is represented in Figure \ref{O43}.
The real mass flow is associated to the motion of some D5 flavor branes along the T-dual circle. 
The branes are rotated until they reach the point $x_3=\alpha'/ \pi r$, that has been named \emph{mirror}
point in \cite{Amariti:2015mva}.
The dual theory is obtained by an HW transition in the IIB setup.
A D3 brane is created each time a D5 brane crosses a non parallel NS brane in this transition.
There are generically two stacks of these D3 branes, one at 
$x_3=0$  and at  $x_3= \pi \alpha'/  r$. This corresponds to a higgsing  of the original gauge symmetry
in the magnetic theory.
Moreover, in the case with real gauge groups, there are also orientifold planes involved: 
an orientifold along $x_3$ in 
type IIA becomes a pair of orientifolds, one at $x_3=0$ and the other at $x_3= \pi \alpha'/  r$ 
\cite{Hanany:2000fq,Hanany:2001iy}. The presence of the extra orientifold at $x_3=\alpha'/ \pi r$  modifies the structure of the HW transition  and of the higgsing.
The 3d limit corresponds to $r \rightarrow 0$. This is a double scaling limit, because this limit
acts on both the geometry and the real mass terms. 

In the magnetic theory the gauge sector at $x_3 = \pi \alpha'/r$ 
is dual to a set of singlets.
By dualizing this sector these singlets have the same quantum charges of the monopoles
of the electric theory, as expected in the Aharony duality and in its generalizations.
This construction has been studied in detail in \cite{Amariti:2015yea,Amariti:2015mva} for many gauge theories, reproducing the field theory expectations.

\subsection{Reducing the index to the partition function}
\label{redind}

In this section we summarize the canonical reduction of the 4d superconformal index to the 3d partition function.
The 4d index is a refined version of the Witten index, defined as
\begin{equation}
I = Tr(-1)^F e^{-\beta H} (pq)^{\frac{\Delta}{2}}
p^{j_1+j_2-R/2} q^{j_1-j_2-R/2}
\prod_{a} u_a^{q_a} 
\end{equation}
where $F$ is the fermionic number, 
$H = \{Q,Q^\dagger\}$, 
$p$ and $q$ are fugacities, $j_1$ and $j_2$ are the third spin components of 
$SU(2)_l \times SU(2)_r = SO(4)$, the isometry of the three sphere and R is the charge of $U(1)_R $.
 The other fugacities $u_a$ are associated to the Cartan of the global symmetry group.
The index is well defined if $p$ and $q$ satisfy the conditions 
\begin{equation}
Im(pq) = 0 \quad |p/q| = 1 \quad |pq| < 1 
\end{equation}
The superconformal index receives contribution only from states with $H = 0$, since
$j_1 \pm j_2 - R/2$
and the internal generators commute with $H$. The superconformal
index can be calculated in two steps. First, one computes the index on single
particle states, also called the single particle index. 
Then the multi-trace index is obtained by the plethystic exponential.
With this procedure one obtains the contributions  of the matter and of the vector multiplets. They correspond to elliptic Gamma functions, defined as
\begin{equation}
\Gamma_e (y;p,q) \equiv \Gamma_e(y) \equiv
\prod_{j,k=0}^{\infty} \frac{1-p^{i+1} q^{j+1}/ y}{1-p^i q^j y}
\end{equation}
where $y$ represent a collective index for the fugacities.
Only gauge invariant states contribute to the index. This imposes an
integration over the holonomy of the gauge group.
The integral is performed
over the fugacity $z_i$, in the Cartan of the gauge group.
We refer the reader to \cite{Dolan:2008qi,Spiridonov:2009za} for details.

The 4d superconformal index reduces to the 3d partition function
computed on the squashed three sphere $S_\mathbf{b}^3$
\cite{Gadde:2011ia,
Dolan:2011rp,Imamura:2011uw}.
The BPS states contributing to the 4d index have to be KK reduced on the circle.
The massless modes in this reduction are the states contributing to the 3d partition function.
In order to perform the KK reduction, it is necessary that all the fugacities
appearing in the index flow to unity. 

As explained in the appendix \ref{doublesc} the KK reduction has to be performed on 
the manifold $S_\mathbf{b}^3 \times \tilde S^1$ instead of $S^3 \times S^1$,
on which the index is naturally defined.
The fugacities are parametrized as functions
of the radius $\tilde r_1$ of $\tilde S^1$
\begin{equation}
p = e^{2 \pi i \tilde r_1 \omega_1}, \quad
q = e^{2 \pi i \tilde r_1 \omega_2}, \quad
u_a = e^{2 \pi i \tilde r_1 \mu_a}, \quad
z_i = e^{2 \pi i \tilde r_1 \sigma_i}
\end{equation}
where $\sigma_i$
is the scalar in the vector multiplet and $\mu_a$ are the scalars in the
background gauge multiplet that become real masses in 3d.
There is also a purely imaginary term associated to the $R$ symmetry.
This term is of the form
 $(\omega_1+\omega_2)R \equiv 2 \omega R$, where $\omega_{1,2}$ are defined in  
 Appendix \ref{sec:appendix}.
This can be absorbed in an imaginary part of the other real masses (this signals the mixing of the $R$ charge 
with the other global symmetries).
In the rest of this section we use a more conventional 3d notation, and we refer 
to the 3d $R$-charge with the mass dimension $\Delta$.
We can use the following identity to perform the limit $\tilde r_1 \rightarrow 0$
\begin{equation}
\label{limit43}
\lim_{
\tilde r_1 \rightarrow 0}
\Gamma_e(e^{2\pi i \tilde r_1 x}; e^{2\pi i \tilde r_1 \omega_1}, 
e^{2 \pi i \tilde r_1 \omega_2})  
=
e^{\frac{i \pi^2}{
6\tilde r_1 \omega_1 \omega_2}
(x-\omega)}
\Gamma_h(x; \omega_1, \omega_2) 
\end{equation}
We will come back to the derivation of this formula in appendix \ref{doublesc}, where we will show how the 
procedure has to be modified to incorporate the double scaling limit we are interested in.
From formula (\ref{limit43}), we can reduce every elliptic gamma function $\Gamma_e$
to a hyperbolic gamma function $\Gamma_h$, which is the building block of the partition function
in 3d. They correctly reduce to the one loop determinants
of the 3d partition function (see appendices \ref{sec:appendix} and  \ref{doublesc}).

The application of this reduction to the identities between the indices of 4d
Seiberg dual phases 
is a standard way to obtain the equality of the partition functions of many 3d dualities.
The equality
of the partition functions follows from the matching of the
corresponding 4d indices. The 4d identities are supported by some conditions on the
fugacities, called balancing conditions.
These conditions become constraint on the real masses in the 3d case. 
The presence of these condition signals the presence of constraints
in field theory, as the anomaly freedom of the R-symmetry or the presence of
superpotential terms.

The final 3d expressions are usually multiplied by a divergent pre-factor,
corresponding to the product of the exponential factors of each field
appearing in (\ref{limit43}).
This pre-factor is proportional to the 4d
gravitational anomalies
and it can be safely removed when we consider
the reduction to 3d of dual theories.

In some cases this procedure requires some modifications. 
For example there can be extra divergences 
in the integrand, related to the presence of a moduli space for the theory on the circle, that spoil the identity.
In this case commuting the 3d limit and the 4d integral gives a divergent result and 
a different reduction scheme has to be performed.
For example this may be the case for theories that develop a moduli space on the circle, as $\mathcal{N}=4$ SYM 
and $SO(N)$ SQCD \cite{Aharony:2013dha}.
A double scaling limit on the real masses and the compactification radius has been proposed.
This is the same double scaling observed in the brane engineering.
In general we observe that this double scaling is necessary whenever the orientifold modifies the
physics of the extra sector at the \emph{mirror} point.
In these cases the orientifold  projects the  unitary gauge  group to a real one also
at $x_3 = \pi \alpha'/r$.
 In the field theory language this is due 
to the periodicity and to the discrete symmetry of the scalar $\sigma$ in the vector multiplet.
This periodicity is lost when we match the partition functions of the
effective 3d pairs obtained by considering the effect of the KK monopoles.
 In this case the correct symmetry breaking pattern of the 3d limit 
 cannot be reproduced in the partition function obtained by the straight reduction of the 4d index
 discussed in this section.
 For this reason it becomes necessary to perform the double scaling limit.
 On the 4d index this limit is obtained as follows: first we consider the effect
 of the real masses as a shift in the fugacities,  that are periodic variables.
 After that we shrink  the radius of the circle and consider the KK reduction on the modes contributing to the index.
This reduction has to be done by considering pairs of $\Gamma_e$, with opposite shifts in the fugacities.
This reproduces the double scaling discussed in the brane scenario. We have to modify 
formula (\ref{limit43}) accordingly. This is discussed in Appendix \ref{doublesc}.
At the end of this limiting procedure we obtain the identities between the partition functions between 3d dual theories
from the 4d ones.
We will give an explicit example of this procedure in the next section, by applying it to the case of $SP(2N)$ SQCD
with adjoint matter.

\section{4d/3d reduction of $SP(2N)$ SQCD with adjoint matter}
\label{sec:reductionSP}
In this section we study the reduction of the 4d duality for symplectic gauge theories 
with adjoint and fundamental matter. We first review the 4d duality of \cite{Leigh:1995qp}.
Then we explain the strategy of the reduction of this duality to 3d, by using the 
argument of  \cite{Amariti:2015yea,Amariti:2015mva}, consisting of the analysis of the brane system,
and we interpret this construction in field theory. 
Moreover, in the field theory analysis, we
modify the construction to make the procedure 
suitable for the reduction
of the 4d index to the 3d partition function.

The main difference in this case is that 
it is not necessary to break the IR theories in decoupled SQCD sectors,
with a polynomial superpotential for the adjoint. 
This is reminiscent of the situation discussed in the unitary case in  \cite{Amariti:2014iza}.

\subsection{The 4d duality}

\begin{itemize}
\item
The electric theory is an $SP(2N)$ gauge theory with 2F fundamentals (F flavors) $Q$  
and a symmetric tensor  $X$ in the adjoint representation of $SP(2N)$. There is a superpotential 
\begin{equation}
\label{Wele}
W = \text{Tr} \, X^{2(k+1)}
\end{equation}
\item
The magnetic theory is an $SP(2\hat N)$ gauge theory, with $\hat N = (2k+1) F-N-2$, with $2F$ 
dual fundamentals $q$ 
and an adjoint $Y$. There are also $2k+1$ gauge singlets $M_j = Q X^j   Q$ in the (anti)symmetric
representation of the flavor symmetry for j odd (even).
 The superpotential of the dual theory is 
\begin{equation}
\label{Wmagn}
W = \text{Tr} \,  Y^{2(k+1)} + \sum_{j=0}^{2k} M_{j}  q Y^{2k-j}  q
\end{equation}
\end{itemize}

\subsection{Brane engineering}

The reduction of this 4d duality to 3d has been discussed in \cite{Amariti:2015yea} in terms of 
D-branes.
Here we summarize this construction and its field theory interpretation.

\begin{figure}
\begin{center}
\includegraphics[width=16cm]{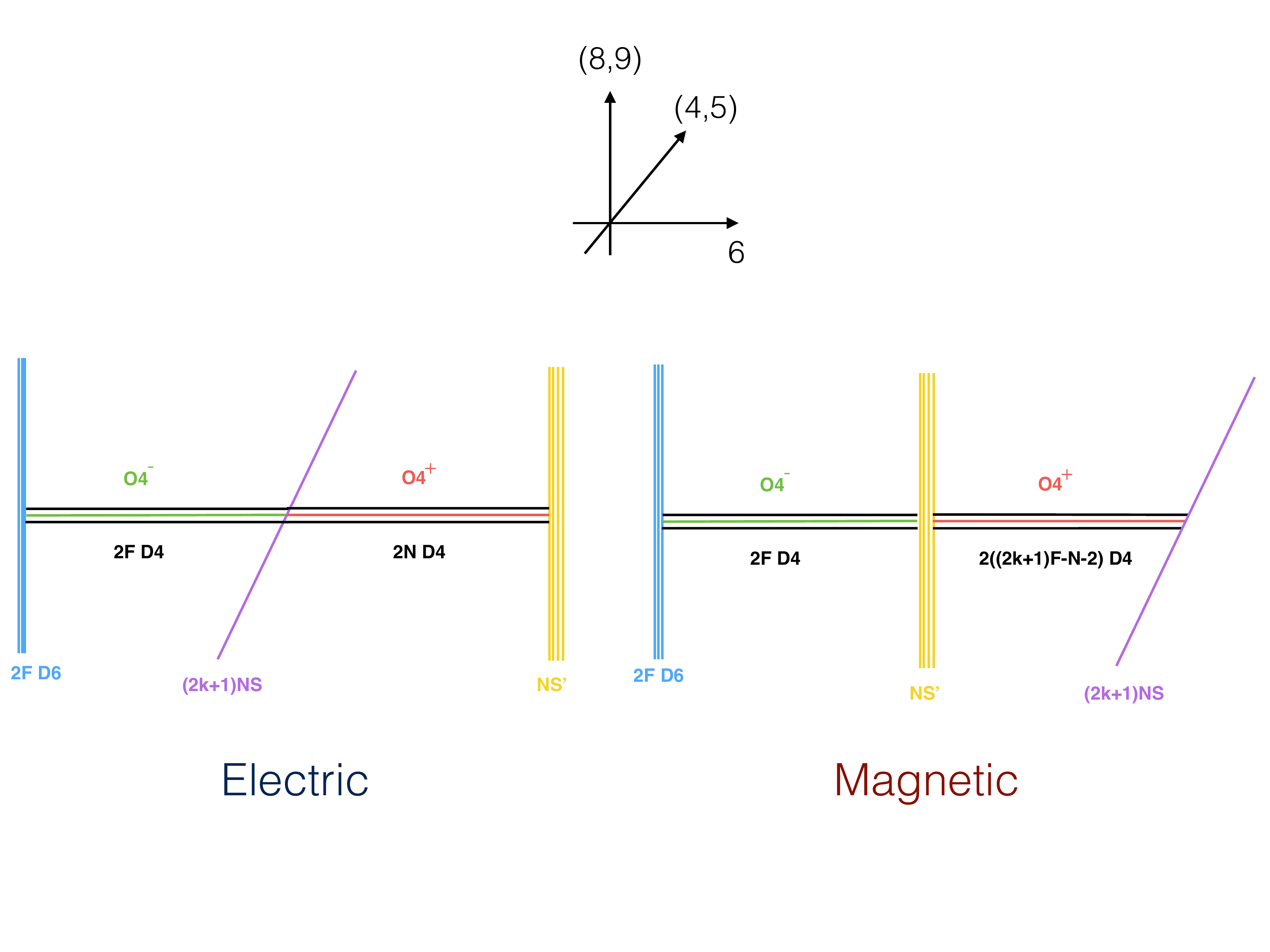}
\caption{In this figure we reproduce the IIA brane setup for the 4d duality between symplectic gauge theories with fundamental and adjoint matter.}
\label{fig:4dDuality}
\end{center}
\end{figure}

Consider a stack of $2N$ D4 branes stretched along the direction 
$x_6$ between an NS' and a stack of  2k+1 NS branes. 
There is also an O4$^{+}$ plane on the stack of D4 branes.
This projects the $U(2N)$  gauge group to $SP(2N)$.
We consider also 2F fields in the fundamental of $SP(2N)$, they are 
associated to a stack of 2F D6 branes.
The branes are extended in ten dimensions as in the table
\begin{equation}
\label{tableD}
\begin{array}{c||cccccccccc}
                 & 0&1&2&3 & 4&5 & 6 & 7 &8&9 \\
\hline
D4             & \times & \times & \times & \times & &       &   \times    &    &        \\
D6             &\times &\times & \times & \times &       &     &&   \times   &      \times     &     \times      \\
NS             &\times & \times & \times & \times &   \times &     \times           &     &    &        \\
NS'            &\times &  \times & \times & \times &       &     &    & &    \times &     \times         \\
O4^{\pm}   &\times &  \times & \times & \times &       &     &    &  &&      
\end{array}
\end{equation}
The dual theory is obtained by an HW transition.
The rank of the dual group is modified 
accordingly to the D-brane charge carried by the orientifold.
 The electric and the magnetic theories are represented in Figure \ref{fig:4dDuality}. This picture is the brane realization of the field theory duality of
\cite{Leigh:1995qp}.

The 4d duality is reduced to 3d by compactifying along $x_3$. The effective 3d duality is captured 
in this brane engineering
by a T-duality along the compact $x_3$ direction.
It leads to a type IIB brane system. In type IIB the NS and NS' branes are left 
unchanged and they are extended
along the compact direction $x_3$. The 
D4 and the D6 branes become D3 and D5 branes respectively and they all sit at the same point on $x_3$. 
This choice is arbitrary at this level, and we can fix these branes at $x_3=0$.
The O4$^+$ plane becomes a pair of (O3$^+$,O3$^+$) planes \cite{Hanany:2000fq,Hanany:2001iy}, one placed at $x_3=0$ and at
the other placed at $x_3=\pi \alpha'/ r$.
By separating the NS branes along the direction $(45)$  we generate a polynomial superpotential 
for the adjoints $X$ and $Y$. The superpotential in the electric and in the magnetic case is  
\begin{equation}
\label{lambda}
W_{ele.}(X) = \sum_{i=1}^{k+1} \lambda_i Tr X^{2i}
\quad \rightarrow \quad
W_{magn.}(Y) = \sum_{i=1}^{k+1} \widetilde \lambda_i Tr Y^{2i}
\end{equation}
where the couplings
$\lambda_i $ and $\tilde \lambda_i  $
are related by the duality.
The superpotential $W_{ele.}(X) $ breaks $SP(2N)$ to $SP(2 r_0) \times \prod U(r_i)$, where
$r_0+r_1+\dots +r_{k} = N$.
On each sector there are BPS and KK monopoles acting as instantons.
In the $SP(2r_0)$ sector a KK superpotential $W_{\eta_0}=\eta_0 Y_0$ is generated, in each $U(r_i)$ sector
a superpotential $W_{\eta_i} = \eta_i Y_i \tilde Y_i$ is generated. The coupling constant  $\eta_i$ corresponds to  $\Lambda_i^{b_i} $, 
where $\Lambda_i$ is the holomorphic scale of the $i$-th sector in the 4d theory and
$b_i$ is the numerator of the 4d NSVZ beta function of the $i$-th sector. 
In the magnetic case the superpotential $W_{magn.}(Y) $
 breaks the $SP(2 \hat N)$ gauge group  as  
$SP(2 \hat  r_0)  \times  \prod U(\hat  r_i)$, where
$\hat  r_0+\hat r_1+\dots  +\hat  r_{2k} = \hat N$.
Also in this case a superpotential for the KK monopoles is generated in each broken gauge sector.

The reduction to 3d is performed by a real mass flow. In the brane interpretation,
on each decoupled SQCD sector,
one rotates two D5 branes along $x_3$  and then performs the HW transition.
Consider for example $2(F+1)$ D5 branes. First we separate the D3 branes along $(45)$ as above.
Then in each sector we rotate two pairs of D5 branes: one D5 clockwise and another D5 counterclockwise,
until they reconnect at $x_3 = \pi \alpha'/ r $.
The electric theory associated to this brane configuration is an $SP(2 r_0)  
\times  \prod U(r_i)$ gauge theory with 2F  fundamentals in the symplectic sector and F flavors in the unitary ones. This theory can  be considered as living at $x_3=0$. The extra matter fields living at $x_3 = \pi \alpha'/r$ are massive and can be integrated out in the IR.

After an HW transition on this brane setup we have 
an $SP(2(F-\tilde r_0-1)) \times \prod_i U(F-\tilde r_i)$ gauge theory 
at $x_3=0$, with 2F fundamentals in the symplectic sector and F flavors in each unitary one. At $x_3=\pi \alpha'/r$ there are also $k$  $U(2)$ theories, each one with two flavors.
These sectors are  separately dual to a set of singlets 
\cite{Aharony:1997bx,Aharony:2013dha}. 
Some of 
these singlets are massive and are integrated out in the IR, some others remain massless and interact with the monopoles of the $U(2r_i)$ gauge groups through an AHW superpotential. These singlets
 survive the compactification, and are interpreted as the electric monopoles, acting as singlets in the dual phase.

\subsection{An alternative  construction}

We can reduce the 4d duality also if we do not turn on the polynomial
superpotential for the adjoint, i.e. with $\lambda_i=\widetilde \lambda_i =0$.
In this case the KK monopoles of the electric and of the magnetic theory are constructed as in \cite{Nii:2014jsa}.
We start from the bare monopoles $Y_0= e^{2 \sigma_1/{g_3^2}+2 i \varphi_1}$
and construct the other monopoles  $Y_j = Tr Y_0 X^{j}$ by dressing 
them with the adjoint matter field, in a way compatible with the 
chiral ring relation, i.e. $j \leq 2k$.

A similar construction has been discussed in \cite{Nii:2014jsa} for the case with adjoint matter and $U(N)$ gauge group. Here the $W_\eta$ superpotential is 
\begin{equation}
W_{\eta}  = \eta Y_0 
\end{equation}
The same conclusions can be reached for the dual case, where $Y_0$ is substituted by
$y_0 = e^{2 \tilde \sigma_1/{\tilde g_3^2}+2 i \tilde \varphi_1}$ 
and $X$ by $Y$.

We can compute the 3d limit by assigning the real masses to two of the fields $Q$ in the 
fundamental representation.
In the dual theory there are two gauge sectors: one sector with $SP(2((2k+1)F-N-1))$ gauge group, 2F fundamentals and an adjoint
and a second sector, due to the higgsing, with $SP(4k)$ gauge group two fundamentals and an adjoint.
This last sector is dual to a set of singlets, and may be used to reconstruct the 
interaction involving the monopole operators in the dual phase
\footnote{We can motivate this claim \emph{a posteriori}: the $SP(4k)$
gauge theory corresponds in 3d to the limiting case of the duality of \cite{Kim:2013cma},
in which the dual theory is described in terms of mesons and monopoles.}.
The $SP(4k)$ sector interacts with the gauge theory at $x_3=0$.
This interaction is an AHW like interaction between the monopoles of the two sectors: the monopoles $y_j = Tr Y^j y_0$ of the 
$SP(2\tilde N)$ gauge theory and the monopoles $\hat Y_j$ of the $SP(4k)$ theory. The dual superpotential is 
\begin{equation}
W = Tr Y^{2(k+1)} + \sum_{j=0}^{2k} M_j q Y^{2k-j} q +
\sum_{j=0}^{2k} y_j \hat Y_{2k-j}
\end{equation}
From the superpotential we can read the charges of the monopoles $\hat Y_j$ , as depicted in (\ref{tablech}): they correspond
to the charges of the monopoles $Y_j$ of the electric theory acting as singlets in the 
magnetic one.
The global charges for the fields involved in this duality are
\begin{equation}
\label{tablech}
\begin{array}{c||ccc}
&SU(2F) & U(1)_A &U(1)_r \\
\hline
\hline
Q&2F&1&\Delta_Q \\
X,Y&1&0&\Delta_X = \Delta_Y = (k+1)^{-1}\\
q &2 \overline F&-1~&\Delta_X- \Delta_Q \\
M_{2j}&F(2F-1)&2&2j \Delta_X +2 \Delta_Q\\
M_{2j+1}&F(2F+1)&2&(2j+1) \Delta_X +2 \Delta_Q\\
Y_j&1&-2F&2F(1-\Delta_Q)-(2N-j)\Delta_X\\
y_j&1&~2F&-2F(1-\Delta_Q)+(2(N+1)+j)\Delta_X
\end{array}
\end{equation}
This is the duality   discussed in \cite{Kim:2013cma}.

\section{The reduction of the index to the partition function}
\label{sec:partition}

In this section we derive the identity between the electric and the magnetic partition functions
of the 3d duality involving $SP(2N)$ gauge groups, fundamental and adjoint matter with 
 superpotentials (\ref{Wele})  and (\ref{Wmagn}).
Let us first summarize the procedure.

First we consider an $SP(4k)$  theory with two fundamentals, an adjoint and
the  superpotential (\ref{Wele}).
This is a limiting case of the 3d duality, namely the dual gauge group 
 is described in terms  of singlets, mesons and monopoles
of the electric theory.
In this case we conjecture the identity between the partition function
of the electric theory and of the dual description.
Observe that this construction is similar to 
the case of  unitary theories with adjoint matter studied in \cite{Amariti:2014iza}.
The main difference is that, in the case at hand here, we have not found an independent derivation
of the 3d identity. Nevertheless, as we will show at the end of this section,  the
relation obtained by dimensional reduction becomes our conjectured identity with an appropriate choice of ranks and parameters.

In the second part of the section we study the reduction to 3d of the duality 
of \cite{Leigh:1995qp}.
First we reduce the 4d identity to a 3d one, for the 
effective duality on the circle.
Then we study the 3d limit on the partition function.
In this case the limit cannot be performed on the 3d identity between the effective theories,
because the compactness of the Coulomb branch is not manifest anymore.

Anyway we show that the 3d limit can be obtained from the 4d index by a double scaling.
It consist of breaking some of the global and local symmetries by shifting some of the flavor and gauge fugacities of the 4d index $t_i \rightarrow t_i e^{i \pi}$. 
Even if the fugacities are periodic on the circle 
we can further consider the effect of this shift in the 3d limit
\footnote{
This limit indeed breaks this periodicity, and the 3d theories have a non compact 
Coulomb branch.}.
The effect  of these shifts turns into the assignation of
real masses, proportional to the inverse radius, to some fundamental fields in the electric theory. In the magnetic phase 
some dual fundamental fields and some mesons acquire a real mass, proportional to the inverse radius.
Moreover there is a
shift involving the gauge fugacities.
It corresponds to the
 higgsing of the gauge symmetry, due to a non-trivial vacuum structure for the scalar in the vector multiplet. 
In this case we observe the emergence of the extra $SP(4k)$ gauge sector on the partition function.
 This sector coincides to the one discussed in the first part of the analysis.
By substituting the identity that we have conjectured for this
$SP(4k)$ sector we obtain the final 3d identity between the partition functions 
of the dual phases of \cite{Kim:2013cma}.

\subsection{A conjecture}

We conjecture a relation between the 3d partition function of an
$SP(4k)$ gauge theory with two fundamentals and an adjoint, with superpotential 
(\ref{Wele}) and a set of singlets, mesons and monopoles. The relation is

\begin{eqnarray}
\label{newrelSP}
Z_{SP(4k)}(\mu;\omega \Delta_X) 
= 
\prod_{j=0}^{2k} \Gamma_h \left( \omega( (j-4k) \Delta_X +2(1-\Delta_{Q})) -2m_A\right)
 \nonumber \\
\prod_{j=0}^{k} \Gamma_h \left(\mu_1+\mu_2+2j \omega \Delta_x\right)
\prod_{j=1}^{k} \prod_{1<a \leq b<2}\Gamma_h \left(\mu_a+\mu_b+(2j-1) \omega \Delta_x\right)
\end{eqnarray}
where we used the conventions of Appendix \ref{sec:appendix}.
We choose the parameter $\mu_a$ in the relation (\ref{newrelSP}) 
as
$\mu_1= \mu_2 = m_A + \omega \Delta_Q$.
In the next section we will use this relation. As a check we will obtain a new relation for 3d partition functions that, in the limiting case, will reduce to (\ref{newrelSP}).

\subsection{The effective duality}

Here we study the reduction of the identity between 
the 4d indices relating the dual phases with symplectic gauge groups,
fundamental and adjoint matter to 3d.
The index of the 4d electric $SP(2N)$ theory with $2(F+1)$ 
flavors and superpotential (\ref{Wele}) is  \cite{Spiridonov:2008zr}
\begin{eqnarray}
\label{Iele}
I_{ele}= &&\frac{(p;p)^{N}(q;q)^{N}}{2^{N} N!} 
\Gamma_e\left( (pq)^{\frac{R_X}{2}} \right)^{N}
\int_{T^{N}} 
\prod_{l=1}^{N} \frac{d z_i}{2 \pi i z_i}
\prod_{i<j} 
\frac{\Gamma_e((pq)^{\frac{R_X}{2}}z_i^{\pm 1}z_j^{\pm 1})}
{\Gamma_e(z_i^{\pm 1}z_j^{\pm 1})}
\nonumber \\
\times &&
\prod_{i=1}^{N}
\frac{\prod_{\alpha=1}^{2(F+1)}  \Gamma_e((pq)^{\frac{R_Q}{2}} s_\alpha z_i^{\pm 1})}
{\Gamma_e(z_{i}^{\pm 2})}
\end{eqnarray}
where $(p;p)$ and $(q;q)$ are the q-Pochhammer symbols.
The fugacity $s_\alpha$ refers to Cartan of the global symmetry of the fundamentals,
while $z_i$ are in the Cartan of the gauge group.

The index of the $SP(2((2k+1)(F+1)-N-2) \equiv SP(2\hat N)$ magnetic theory with $2(F+1)$ dual quarks,
an adjoint $Y$, the $k+1$ antisymmetric  mesons $M_{2j}$ and the $k+1$
symmetric mesons $M_{2j-1}$ and the superpotential (\ref{Wmagn})
is  \cite{Spiridonov:2008zr} 
\begin{eqnarray}
\label{Imagn}
&&I_{magn}
= \frac{(p;p)^{\hat N}(q;q)^{\hat N}}{2^{\hat N} \hat N!}
\prod_{j=0}^{k} \prod_{\alpha<\beta}^{2(F+1)}
\Gamma_e((pq)^{\frac{R_{M_{2j}}}{2}} s_\alpha s_{\beta}))
\prod_{j=1}^{k} \prod_{\alpha \leq \beta}^{2(F+1)}
\Gamma_e((pq)^{\frac{R_{M_{2j-1}}}{2}} s_\alpha s_{\beta}))
\nonumber \\
&&
\Gamma_e\left( (pq)^{\frac{R_Y}{2}} \right)^{\hat N}
\int_{T^{\hat N}} 
\prod_{i=1}^{\hat N} \frac{d z_i}{2 \pi i z_i}
\frac{
\prod_{\alpha=1}^{2(F+1)} 
\Gamma_e((pq)^{\frac{R_X-R_Q}{2}} s_\alpha z_i^{\pm 1})}
{\Gamma_e(z_{i}^{\pm 2})}
\prod_{i<j} 
\frac{\Gamma_e((pq)^{\frac{R_Y}{2}}z_i^{\pm 1}z_j^{\pm 1})}
{\Gamma_e(z_i^{\pm 1}z_j^{\pm 1})}
\nonumber \\
\end{eqnarray}
The identity between (\ref{Iele}) and (\ref{Imagn}) is guaranteed if the 
fugacities satisfy the following balancing condition \cite{Spiridonov:2009za}
\begin{equation}
\label{eq:balancing4d}
(pq)^{2 R_X (N+1)} \prod_{\alpha=1}^{2(F+1)} s_\alpha = (pq)^{2(F+1)}
\end{equation}
Here we consider the reduction of the identity  between (\ref{Iele}) and (\ref{Imagn})
 to a 3d identity between the partition functions
computed on  $S_\mathbf{b}^3$. We follow the strategy explained in section \ref{redind}.
The partition function of the electric theory,
with the conventions of Appendix \ref{sec:appendix},  becomes 
\footnote{We modify the label of the R-charges form $R$ to $\Delta$ when considering the 3d theories. 
Here it is a change of variables because the
constraint on the R-charges, imposed by the anomalies in 4d, are imposed in 3d, on the circle,
by the $W_\eta$ superpotential.}
\begin{equation}
\label{eleSPs}
Z_{ele} = Z_{SP(2N)}(\mu;\Delta_X)
\end{equation}
and the 4d balancing condition 
(\ref{eq:balancing4d}) becomes the 3d balancing condition
\begin{equation}
\label{eq:balancing3d}
\sum_{a=1}^{2(F +1)} \mu_a = 2\omega \left(F+1-(N+1)\Delta_X  \right)
\end{equation}
This last condition constraints the real mass parameters and the
R-charges of the theory, breaking the otherwise generated $U(1)_A$ symmetry.
This reflects in the superpotential $W_\eta$ in the field theory analysis.
The partition function of the dual model becomes
\begin{eqnarray}
\label{duakm}
Z_{magn} &=&  
\prod_{j=0}^{k} \prod_{a<b}^{2(F+1)}\Gamma_h \left(\mu_a+\mu_b+2j \omega \Delta_x\right)
\prod_{j=1}^{k} \prod_{a \leq b}^{2(F+1)}\Gamma_h \left(\mu_a+\mu_b+(2j-1) \omega \Delta_x\right)
\nonumber \\
&\times&
Z_{SP(2 \hat N)}(\omega \Delta_X -  \mu;\omega \Delta_Y)
\end{eqnarray}
In both the formulas (\ref{eleSPs}) and (\ref{duakm}) 
there is also a divergent pre-factor. We checked that it coincides in both the cases.

\subsection{The 3d limit}

The last step consists of reducing the identity between the theories on the circle to an
identity for the pure 3d duality.
As observed in the brane analysis this requires a double scaling limit in the reduction of
the 4d duality. 
This is necessary because at the \emph{mirror} point an $O3 $ orientifold projects 
an $U(N)^2$ gauge algebra to 
$SP(2N)$, $SO(2N)$ or $SO(2N+1)$ .
In the field theory analysis this projection reflects in the fact that,
if  some components of the periodic scalar $\sigma$
in the vector multiplet are reconnected at the \emph{mirror} point,  the unitary group
becomes a real one.

On the partition function the scalar $\sigma$ is not periodic but it is integrated over the 
entire real axis, because the duality on the circle is considered as an effective 3d duality.
The role of the finite circle and, as a consequence, of the extra orientifold at the \emph{mirror} point, are not evident in this picture.

The double scaling limit, discussed in the brane analysis, 
has to be rephrased here, in the reduction of the 
4d index.
This is done by pairing the elliptic Gamma functions  of the fields that acquire, in the 3d limit, a
large opposite real mass. 
Then we shift the fugacities associated to the 3d real masses by an amount of $e^{i \pi}$.
The same procedure has to be applied both to the electric and to the magnetic theory.

The elliptic Gamma functions are then expanded in KK modes, as explained in Appendix \ref{doublesc}.
There are three possible terms. 
First there are terms that are not shifted: these modes are treated as above, i.e. formula (\ref{limit43})
can applied on these modes.
Then there are terms on which the shift of the fugacity is proportional to $e^{2 \pi i}=1$.
Also these modes are treated again with formula (\ref{limit43}). 
Eventually there are modes with a shift proportional to $e^{i \pi}$.
In this case we have to use the results of Appendix \ref{doublesc}, 
pairing these modes with the ones with opposite 
shift. 
  Observe that the shifts become real mass terms 
  in the field theory analysis, proportional to the inverse radius
  $\tilde r_1$ (see the appendices \ref{sec:appendix}
  and \ref{doublesc} for the conventions on the geometry).
  Here we shift $s_{2F+1} \rightarrow s_{2F+1} e^{i \pi}$  and
  $s_{2F+2} \rightarrow s_{2F+2} e^{-i \pi}$  
  in the electric 4d index. 
The real masses associated to the shifts, in the 3d electric theory, become 
\begin{equation}
\label{elemu}
\mu_a = 
\left\{
\begin{array}{rl}
     m_A + m_a+ \omega \Delta_Q   ~ &~~~~~a=1,\dots,2F   \\
-F m_A + \frac{1}{2\tilde r_1}  + \omega \Delta_{Q_1} &~~~~~ a=2F+1\\
-F m_A - \frac{1}{2 \tilde r_1}  + \omega \Delta_{Q_1} &~~~~~ a= 2F+2       
\end{array}
\right.
\end{equation}
where the finite shifts in the real masses are due to the 
non abelian symmetry. These finite terms represent the axial symmetry 
that is generated in the 3d theory.
In the dual phase the shifts and the masses are assigned consistently with the
global symmetries. 
In the magnetic theory we consider also the shift $z_{i} \rightarrow z_{i} e^{i \pi}$
for $i=\tilde N + 1,\dots, \tilde N + 2 k\equiv \hat N$.
 This breaks the dual gauge group to 
$SP(2 \tilde N) \times SP(4k)$.
By computing the double scaling limit we arrive at the 3d relation
\begin{eqnarray}
\label{intermediate}
Z_{SP(2N)}( \mu;\omega \Delta_X) 
&=& 
Z_{SP(2 \tilde N)} \left(
\omega \Delta_X - \mu;\omega \Delta_Y \right)
\prod_{j=0}^{k}\prod_{a <b}^{2F}
\Gamma_h(\mu_a+\mu_b+ 2 j \omega \Delta_X)
  \\
&\times&
\prod_{j=1}^{k}\prod_{a\leq b}^{2F}
\Gamma_h(\mu_a+\mu_b+(2j-1) \omega \Delta_X)
\prod_{j=0}^{k} \Gamma_h (2 j \omega \Delta_X-\nu_1-\nu_2)
\nonumber  \\
&\times &
\prod_{j=1}^{k} \prod_{a=1,2} \Gamma_h ((2j+1) \omega \Delta_X-\nu_a-\nu_b)
\quad
Z_{SP(4k)}(\nu; \omega \Delta Y)
\nonumber
\end{eqnarray}
where $\mu$  refers to the first $2F$ entries in
(\ref{elemu}).
The term $Z_{SP(4k)}(\nu; \omega \Delta Y)$ in (\ref{intermediate}) 
refers to an
$SP(4k)$ sector with an adjoint with W~=~Tr~$Y^{2(k+1)}$ and two fundamentals,
where $\nu$ is a two component vector 
\begin{equation}
\nu_{i} = m_A F + \omega(\Delta_X-\Delta_{Q_1}) \quad i=1,2
\end{equation}
The value of $\Delta_{Q_1}$ is obtained from the balancing condition (\ref{eq:balancing3d}).
We have $\Delta_{Q_1} = 1+F(1-\Delta_Q)-(N+1) \Delta_X $.
The partition function can be computed from the relation conjectured in (\ref{newrelSP}).
By substituting the expression (\ref{newrelSP}) in (\ref{intermediate}), 
and
by using the identity $\Gamma_h(2\omega-x)\Gamma_h(x)=1$, we 
have
\begin{eqnarray}
\label{final}
&&
Z_{SP(2N)}(\mu;\omega \Delta_X) = 
Z_{SP(2\tilde N)} \left(
\omega \Delta_X-\mu;\omega \Delta_Y \right)
\prod_{j=1}^{k}\prod_{a\leq b}^{2F}
\Gamma_h(\mu_a+\mu_b+(2j-1) \omega \Delta_X)
\nonumber \\
&&
\prod_{j=0}^{k}\prod_{a <b}^{2F}
\Gamma_h(\mu_a+\mu_b+ 2 j \omega \Delta_X)
\prod_{j=0}^{2k}
\Gamma_h ( \omega (2F (1\!-\!\Delta_Q)+(j\!-\! 2N) \Delta_X)\!-\!2 m_A F)
\end{eqnarray}
This is the main result of our paper, it represents the relation between the electric and the magnetic partition functions for the dual phases of
\cite{Kim:2013cma}, in the case of symplectic gauge groups.
Observe that the relation (\ref{final})  reduces to the one studied in \cite{Willett:2011gp} for the case $k=1$,
where the adjoint field is massive and the standard duality of \cite{Aharony:1997gp} is obtained.
Note also that  (\ref{final})  reduces to 
the conjectured expression (\ref{newrelSP}) for  $N=2k$ and $F=1$.

\section{SP(2N) SQCD with an antisymmetric tensor}
\label{sec:antisymmetric}

In this section we discuss the case of $SP(2N)$ with $2F$ fundamentals and a traceless antisymmetric tensor.
The 4d duality has been studied in \cite{Intriligator:1995ff}
\begin{itemize}
\item
The electric theory is an $SP(2N)$ gauge theory with $2F$  fundamentals $Q$ 
and a traceless antisymmetric $A$. There is a superpotential 
\begin{equation}
\label{Wele2}
W = Tr A^{k+1}
\end{equation}
\item
The magnetic theory is an $SP(2\hat N)$ gauge theory, with $\hat N = (F-2) k-N$, with $2F$ fundamentals $q$ and a traceless 
antisymmetric  tensor 
$a$. There are also $k$ gauge singlets $M_j = Q A^j   Q$.
 The superpotential is
\begin{equation}
\label{Wmagn2}
W = Tr a^{k+1} + \sum_{j=0}^{k-1} M_{k-j-1}  q a^j  q
\end{equation}
\end{itemize}
In this case one can break the $SP(2N)$ theory in $\prod SP(2r_i)$ and the dual 
$SP(2\hat N)$ theory in $\prod SP(2 \hat r_i)$ and reduce in each sector.
This case has been discussed in details in the brane scenario in \cite{Amariti:2015mva}.
In alternative one can consider the gauge groups as unbroken and perform the reduction 
as done above, in the case with the adjoint.
The main difference in this case is that, when computing the 3d limit,  the extra sector in the magnetic theory, 
placed at $x_3 = \pi\alpha'/ r$, consists of a set of singlets.
In this case one can follow the reduction from the 4d  index to 3d the partition function
for the effective duality and compute the 3d limit of this identity, 
as done in \cite{Aharony:2013dha} for $SP(2N)$ SQCD.
We leave the details of the derivation to the interested reader
and here we just state the results.

By reducing the identity on the 4d index for the dual pair
of \cite{Intriligator:1995ff} we obtain the relation
\begin{equation}
\label{SPAS}
Z_{SP(2N)}^{(A)}(\mu,\omega \Delta_A)
=
\prod_{j=0}^{k-1} \prod_{\alpha<\beta} 
\Gamma_h(\mu_\alpha+\mu_\beta+j \omega \Delta_A)
Z_{SP(2 \hat N)}^{(a)}(\omega \Delta_A-\mu,\omega \Delta_a) 
\end{equation}
where the partition function  
for an symplectic gauge theory with fundamentals
and a traceless antisymmetric tensor has been 
defined in the appendix \ref{sec:appendix}.
The dual rank is $\hat N = k(F-2)-N$ and  $\Delta_A = \Delta_a = 2(k+1)^{-1}$.
The relation (\ref{SPAS}) represents the matching of the partition 
functions for the effective duality, on the circle.
The balancing condition is 
\begin{equation}
\label{balaSPA}
\sum_{\alpha=1}^{2F} \mu_\alpha + 2(N-1) \omega \Delta_A = 2 (F-2) \omega
\end{equation}
In the limiting case, $\hat N = 0$, obtained by assigning $N=k$ and $F=3$ the relation (\ref{SPAS})
corresponds to the relation (5.3.7) of \cite{VdB} and the balancing condition 
(\ref{balaSPA}) coincides with  (5.3.5) of \cite{VdB}.

The 3d duality is obtained by considering $2F+2$ flavors, and 
assigning a large real mass to the last two.
We obtain the relation
\begin{eqnarray}
\label{SPA}
&&
Z_{SP(2N)}^{(A)}(\mu,\omega \Delta_A)
=
\prod_{j=0}^{k-1} \prod_{\alpha<\beta} \Gamma_h(\mu_\alpha
+\mu_\beta+j \omega \Delta_A)
Z_{SP(2 \tilde N)}^{(a)}(\omega \Delta_A-\mu,\omega \Delta_a) 
\nonumber \\
&&
\prod_{j=0}^{k-1} 
\Gamma_h( \omega  \Delta _A (j-2 (N-1))-2 \omega  \left(1-F \left(1-\Delta _Q\right)\right)-2 F m_A )
\end{eqnarray}
where the last line corresponds to the contribution of the 
monopoles.
In this case the rank of the dual gauge group is  $\tilde N = k(F-1)-N$.

Observe that we can also study the case of the antisymmetric tensor
studied in \cite{Kapustin:2011vz}. In this case the antisymmetric can 
be decomposed in a traceless part and a singlet. The effect 
consists of multiplying on the LHS (RHS) side of the identities 
(\ref{SPAS}) and (\ref{SPA}) 
the factor $\Gamma_h(\omega \Delta_A)$ ($\Gamma_h(\omega \Delta_a)$).

\section{Counclusions}
\label{sec:conclusions}

In this paper we studied the reduction of 4d Seiberg duality for symplectic SQCD with adjoint matter
to 3d. We discussed the field theory aspects of the reduction, reproducing the results of
 \cite{Amariti:2015mva}, derived from the brane perspective.
The main result of our paper concerns 
the reduction of the 4d index to the 3d partition function:
we obtained new mathematical identities and reproduced the field theory expectations.

Our derivation is based on a double scaling limit on the radius and on the real masses.
Moreover we conjectured the relation (\ref{newrelSP}), necessary for the derivation of the results. 
This conjecture has a physical justification but it deserves a deeper mathematical analysis. 
We leave this as open questions for future investigations.

The analysis can be extended to 4d dualities involving orthogonal gauge groups. In these cases  
the theories on the circle have a moduli space that makes the reduction more subtle.
For example the 3d partition functions obtained from the 4d indices are divergent 
\cite{Aharony:2013dha}. One can still obtain the 3d identities for the pure 3d dual theories
by using the double scaling discussed above.
One can also study unitary cases with tensor matter: in these cases the field theory aspects of the reduction can be understood from the brane construction and a similar analysis on the reduction
4d index is required.

\section*{Acknowledgments}
A.A. is grateful to Claudius Klare for early collaboration on this project
and for carefully reading the draft. 
   A.A. is funded by the European Research Council
    ERC-2012-ADG\_20120216 and
    acknowledges support by ANR grant
    13-bs05-0001. A. A. would like to thank  University of Milano-Bicocca for hospitality during various stages of this work.

\appendix
\section{The partition function of SP(2N) gauge theories }
\label{sec:appendix}

In this appendix we provide some mathematical definitions concerning the
3d partition function of symplectic SQCD with tensor matter.
The squashed 3-sphere $S^3_\mathbf{b}$ is defined by the equation
$\mathbf{b}^2(x_1^2+x_2^2)+\mathbf{b}^{-2}(x_3^2+x_4^2)=1$,
where $x_i$ are real coordinates and $\mathbf{b}$ is the real squashing parameter.
The partition function on  $S_\mathbf{b}^3$ for a gauge group $SP(2N)$  
is a matrix integral over the Cartan of the gauge group,
parameterized by the scalar $\sigma$ in the $\mathcal{N}=2$ vector multiplet \cite{Jafferis:2010un,Hama:2010av,Hama:2011ea}.
For $2F$ fundamental and one  adjoint fields  the matrix integral is
\begin{eqnarray}
\label{eq:Zdef} 
Z_{SP(2N)} \big( \mu; \tau \big) =
&&\frac{\Gamma_h(\tau^n)}{\sqrt{-\omega_1 \omega_2}^N 2^N N!} 
\int \prod_{i=1}^{N} d \sigma_i 
\frac{\Gamma_h(\tau \pm 2 \sigma_i)}{\Gamma_h(\pm 2\sigma_i )} 
\nonumber \\
\times 
&&
\prod_{i=1}^{N}
\prod_{a=1}^{2F} \Gamma_h(\mu_a \pm \sigma_i)
  \prod_{1 \le i < j \le N } \!\! \frac{\Gamma_h(\tau \pm \sigma_i \pm \sigma_j)}{\Gamma_h(\pm \sigma_i \pm \sigma_j)}  \nonumber \\
 \end{eqnarray}
where $\omega_1=i \mathbf{b}$ and $\omega_2=i \mathbf{b}^{-1}$.
The function $\Gamma_h$ is the hyperbolic Gamma function. For Im$(\omega_2/\omega_1)>0$
 it can be written as \cite{VdB}
\begin{equation}
\label{eq:Gammahvbd}
\Gamma_h(x;\omega_1,\omega_2) \equiv
\Gamma_h(x)\equiv 
e^{
\frac{i \pi}{2 \omega_1 \omega_2}
((x-\omega)^2 - \frac{\omega_1^2+\omega_2^2}{12})}
\prod_{j=0}^{\infty} 
\frac
{1-e^{\frac{2 \pi i}{\omega_1}(\omega_2-x)} e^{\frac{2 \pi i \omega_2 j}{\omega_1}}}
{1-e^{-\frac{2 \pi i}{\omega_2} x} e^{-\frac{2 \pi i \omega_1 j}{\omega_2}}}
\end{equation}
We also introduced the notation $\Gamma_h(\pm x) \equiv \Gamma_h(x)\Gamma_h(-x)$.
The hyperbolic Gamma functions are the $1$-loop contributions of the various multiplets,
their arguments represent the weights of the scalars in the vector multiplet of the local and the global symmetries.
More precisely, $\pm \sigma_i  \pm \sigma_j$ (with $i<j$)
and $\pm 2 \sigma_i$ are the
non-zero weights\footnote{There are also $n$ zero weights, that have to be considered 
in the matter field but not in the vector multiplet, because they
cancel against the contribution of the measure. The zero weights 
are summarized in the term $\Gamma_h(\tau)^n$ in (\ref{eq:Zdef}).} of the adjoint representation while $\pm \sigma_i$ are the ones of the fundamental.
Similarly $\tau$ and the vector $\mu$ parameterize the weights under the global symmetries
for the adjoint and for the fundamental matter multiplets.

The partition function for an $SP(2n)$ gauge theory with 2F fundamentals
and a traceless antisymmetric tensor is 
\begin{equation}
Z_{SP(2n)}^{(A)}(\mu,\tau_A) = 
\frac{\Gamma_h(\tau_A)^{n-1}}{\sqrt{-\omega_1 \omega_2}^n 2^n n!} 
\int \prod_{i=1}^{n} d \sigma_i 
\prod_{i<j} 
\frac{\Gamma_h(\pm \sigma_i \pm \sigma_j + \tau_A)}
{\Gamma_h(\pm \sigma_i \pm \sigma_j )}
\prod_{i=1}^{n} 
\frac{\prod_{\alpha =1}^{2F} \Gamma_h(\mu_\alpha \pm \sigma_i)}
{\Gamma_h(\pm 2 \sigma_i)}
\end{equation}
where in this case the non-zero weights of the antisymmetric tensor are
$\pm \sigma_i  \pm \sigma_j$ (with $i<j$). There are also $n-1$ zero weights.

\section{The double scaling limit}
\label{doublesc}

In this appendix we discuss the double scaling limit in the reduction of the 4d
index to the 3d partition function. 
First we review the standard reduction, as explained in 
\cite{Aharony:2013dha}. Then we show how to modify this
procedure when the double scaling limit is considered.

The 4d superconformal index is computed on $S^3 \times S^1$,
where one refers to the radius of the 3-sphere and of the circle as
$r_3$ and $r_1$ respectively.
Alternatively one can compute the index on the product space 
$S_\mathbf{b}^3 \times \tilde S_1$ 
\cite{Agarwal:2012hs,Aharony:2013dha,Assel:2014paa,Closset:2014uda,Assel:2015nca}.
In this case  $S_\mathbf{b}^3$ is the squashed three sphere discussed
above and $\tilde S_1$ is a different circle, with radius $\tilde r_1 = 2(\mathbf{b}+\mathbf{b}^{-1})^{-1} r_1$.

We study the reduction of the 4d index $I(u;p;q)$ of a free field to the 3d partition function 
$Z(m;\omega_1;\omega_2)$
as a KK decomposition on the circle $\tilde S^1$.
The relation between the fugacities and the real mass parameters is 
\begin{equation}
p = e^{2 \pi i \tilde r_1 \omega_1}, \quad
q = e^{2 \pi i \tilde r_1  \omega_2}, \quad
u = e^{2 \pi i \tilde r_1  m}
\end{equation}
Formally we can write the decomposition as
\begin{eqnarray}
\label{KKred}
I(u;p,q) = \prod_{n=-\infty}^{\infty} Z\big(m+\frac{n}{\tilde r_1};\omega_1,\omega_2 \big)
\end{eqnarray}
Then one has to define a more proper form of the 4d index  for the reduction
to 3d.
It consists of taking a different normalization. This is required because the 
here we refer to the  SUSY partition
function on $S_b^3 \times \tilde S^1$.
The two expressions for the 4d index differ by a normalization $I_0(m)$, 
\begin{equation}
\label{casimir}
I(u;p;q) = e^{I_0(m)} \Gamma_e (u;p,q)
\end{equation}
The normalization factor $I_0(m)$ has been obtained by \cite{Kim:2012ava}, as the solution of the equation
\begin{equation}
\frac{I_0(m)}{\tilde r_1} = \frac{1}{4} \left(
r^{-1} \partial_r(r \Gamma_0 (u;p;q)\right)|_{r=0}
\end{equation}
where $\Gamma_0(u;p;q) = (u-pq/u)/((1-p)(1-q))$ is the single particle index.
In the  case at hand $\Gamma_0$ has a simple pole at
$r=0$ and there are not constant terms in $r$. The solution gives
\begin{equation}
\label{io}
I_0(m) = \frac{i \pi \tilde r_1 (m-\omega)(2m(m-2 \omega)+\omega_1 \omega_2}{6 \omega_1 \omega_2}
\end{equation}
The second step in the derivation consists of studying the partition function
 $Z = \Gamma_h$.
Here we restrict to a free field. 
The infinite product in \ref{KKred}, involving $\Gamma_h$, is divergent and it needs to be properly regularized.
For example by choosing the  zeta-function  regularization, $\sum n^s = \zeta(-s)$, one obtains 
 \begin{equation}
 \label{KKsumreg}
e^{I_0(m)} \Gamma_e(u;p,q) =
e^{-\Delta}
\prod_{n=-\infty}^{\infty}
e^{-sign(n) \frac{i \pi}{2 \omega_1 \omega_2} 
\left(\left(m+\frac{n}{\tilde r_1}-\omega\right)^2 - \frac{\omega_1^2+\omega_2^2}{12}\right)}
\Gamma_h\left(m+\frac{n}{\tilde r_1};\omega_1;\omega_2\right)
  \end{equation}
where
\begin{equation}
\Delta = \frac{i \pi  \left(-2 \omega  (6 m \tilde r_1+1)+2 m (3 m \tilde r_1+1)+\tilde r_1 \left(\omega _1^2+3 \omega _2 \omega _1+\omega _2^2\right)\right)}{12 \tilde r_1 \omega _1 \omega _2}
\end{equation}
The relation (\ref{KKsumreg}) involves regular functions. The next step consists of computing the limit $\tilde r_1 \rightarrow 0$. It is performed by observing that only the term with $n=0$ survives in the product.
Finally one obtains the relation (\ref{limit43}).

In the rest of this appendix we discuss a different limit, that we have used in the main paper.
This consists of a double scaling limit, and it is the limit discussed in the brane analysis.
In the partition function we shift two fugacities $u^{(1)}$ and $u^{(2)}$ as  $u^{(1)} \rightarrow u^{(1)} e^{ i \pi}$
and $u^{(2)} \rightarrow u^{(2)} e^{- i \pi}$. 
This reflects in the shifts $m^{(1)} \rightarrow m^{(1)}   + 1/2 \tilde r_1$ 
and $m^{(2)} \rightarrow m^{(2)}   - 1/2 \tilde r_1$ on the associated real masses.
At this point we perform the double scaling limit: while shrinking the radius 
the real masses become large. 
In the brane analysis this limit corresponds to  
first rotating the pairs of D5 branes at the \emph{mirror} point,  
and then by sending the radius of the T-dual circle to infinity. 
This translates here in performing this limit on pairs
of elliptic gamma functions, with opposite shifts in the fugacities
\footnote{Observe that it is not necessary to make the opposite shift on elliptic gamma functions with an opposite fugacity.}.

In the rest of this section we show how to modify the standard reduction in this case.  
Consider the following relation
\begin{eqnarray}
\label{prodG}
I(u^{(1)} e^{i \pi \tilde r_1})  I(u^{(2)} e^{- i \pi \tilde r_1})  
= \prod_{n=-\infty}^{\infty} 
Z(m^{(1)}+\frac{n }{\tilde r_1} + \frac{1}{2 \tilde r_1})
Z(m^{(2)}+\frac{n }{\tilde r_1} - \frac{1}{2 \tilde r_1})
\end{eqnarray}
where for simplicity we omitted the $p,q$ and $\omega_i$ dependence.
The equation (\ref{casimir}) here becomes
\begin{equation}
\label{casimir2}
I(u^{(1)} e^{i \pi \tilde r})  I(u^{(2)} e^{- i \pi \tilde r_1})  
= e^{I_0(m^{(1)})+I_0(m^{(2)})} 
  \Gamma_e (u^{(1)} e^{i \pi \tilde r_1})  
    \Gamma_e(u^{(2)} e^{- i \pi \tilde r_1})  
  \end{equation}
where  $I_0(m)$ coincides with the one obtained in (\ref{io}).
The reason is that the linear terms in $\Gamma_0$ 
generated because of the shifts $n \rightarrow n \pm \frac{1}{2}$
cancel in the product.
Next we have to regularize the divergence in the product 
hyperbolic gamma functions in the RHS of (\ref{prodG}).
We use again the zeta-function regularization as before.
After the regularization we compute the limit and obtain
\begin{equation}
\label{doublesca}
\lim_{\tilde r \rightarrow 0} 
\Gamma_e(u_1 e^{i \pi} ;p;q) \Gamma_e(u_2 e^{-i \pi} ;p;q) =
e^{-\frac{\pi i ( m^{(1)} + m^{(2)}  - 2 \omega) }
{6 \tilde r_1 \omega_1 \omega_2 }}
\end{equation}
where we can turn on the $R$ charge dependence by assigning an imaginary part to the
real masses $m^{(i)}$.
The relation (\ref{doublesca}) can be modified for generic fugacities and R charges
and it can be written in the same form of (\ref{limit43}).
It becomes 
\begin{equation}
\label{doublesca}
\lim_{\tilde r \rightarrow 0} 
\Gamma_e(u_1 e^{i \pi} ;p;q) \Gamma_e(u_2 e^{-i \pi} ;p;q) =
e^{-\frac{\pi i (\sum_a m_a^{(1)} e_a + \sum_b m_b^{(2)} e_b  -  \omega(2-R^{(1)} -R^{(2)}) }
{6 \tilde r_1 \omega_1 \omega_2 }}
\end{equation}

\bibliographystyle{JHEP}
\bibliography{BibFile}

\end{document}